\documentclass[aps,prl ,showpacs,twocolumn,groupedaddress,superscriptaddress,nobibnotes
]{revtex4-2}

\usepackage[english]{babel}

\usepackage[dvipsnames]{xcolor}
\usepackage{graphicx}
\usepackage{dcolumn}
\usepackage{bm}
\usepackage{upgreek}
\usepackage[normalem]{ulem}

\usepackage{amssymb}
\usepackage{amsmath}

\usepackage{xr}
\makeatletter
\newcommand*{\addFileDependency}[1]{
  \typeout{(#1)}
  \@addtofilelist{#1}
  \IfFileExists{#1}{}{\typeout{No file #1.}}
}
\makeatother

\begin{document}


\title{Magnetic-field switching of exciton-magnon coupling in LiNiPO$_4$ }

\author{Bei Sun}
\affiliation{Institute for Solid State Physics, The University of Tokyo, Chiba, 277-8581, Japan}

\author{Zhuo Yang}
\affiliation{Institute for Solid State Physics, The University of Tokyo, Chiba, 277-8581, Japan}

\author{Julian Shibuya}
\altaffiliation{Present address: School of Materials and Chemical Technology, Institute of Science Tokyo, Kanagawa, 226-8501, Japan}
\affiliation{Department of Materials Science, Osaka Prefecture University, Osaka 599-8531, Japan }

\author{Koichi Kindo}
\affiliation{Institute for Solid State Physics, The University of Tokyo, Chiba, 277-8581, Japan}

\author{Kenta Kimura}
\affiliation{Department of Materials Science, Osaka Metropolitan University, Osaka 599-8531, Japan}

\author{Atsuhiko Miyata}
\email{a-miyata@issp.u-tokyo.ac.jp}
\affiliation{Institute for Solid State Physics, The University of Tokyo, Chiba, 277-8581, Japan}


\newcommand{\Joey}[1]{{\color[RGB]{26,111,223}{#1}}}
\newcommand{\Joeyc}[1]{{\color[RGB]{241,64,64}{[Joey: {\it #1}\,]}}}
\newcommand{\Miyata}[1]{{\color[RGB]{26,111,223}{#1}}}

\clearpage

\begin{abstract}

Exciton–magnon transitions provide a fundamental optical fingerprint of coupled excitonic and magnetic excitations in antiferromagnets.  
However, controlling such coupled excitations by external fields remains a key challenge.
Here we report the temperature and magnetic-field evolution of exciton-magnon coupling in the magnetoelectric antiferromagnet LiNiPO$_4$ using pulsed magnetic fields up to 50 T.
The magnon sideband intensity exhibits sharp switching across field-induced magnetic phases, with strong suppression in plateau phases and enhancement in canted spin states. This behavior is attributed to the interplay between the thermal magnon population and the spin-dependent optical transition matrix element. These results demonstrate that magnetic-field control of spin degrees of freedom enables selective switching of exciton-magnon coupling in antiferromagnets.

\end{abstract}

\maketitle

\clearpage


Excitons—elementary quasiparticles consisting of bound electron–hole pairs in semiconductors and insulators—play a crucial role in mediating light-matter interactions \cite{HaugKoch1993,YuCardona2010,ChemlaShah2001,Koch2006}. The study of excitonic states provides fundamental insight into the microscopic optical processes and enables the development of advanced optoelectronic and quantum technologies \cite{HaugKoch1993,YuCardona2010,ChemlaShah2001,Koch2006}. Recently, the emergence of diverse antiferromagnetic systems, e.g., van der Waals materials, 
 has triggered renewed interest in the interplay between excitonic excitations and spin degrees of freedom \cite{Bae2022NatureCrSBr,Dirnberger2026CrSBr,Klaproth2023PRL,Kang2020Nature,Song2024NiPS3,Son2022NiI2}. These systems host unique excitonic phenomena, including magnetic excitons and Zhang–Rice excitons, where optical responses are strongly influenced by the underlying spin structure \cite{Bae2022NatureCrSBr,Dirnberger2026CrSBr,Klaproth2023PRL,Kang2020Nature,Song2024NiPS3,Son2022NiI2}. 

A key manifestation of such coupling is the appearance of exciton–magnon transitions (i.e., magnon sidebands) \cite{TanabeMoriyaSugano1965,Allen1966PRL,Sell1967,MeltzerMnF2}. In such processes, an exciton is created simultaneously with the emission or absorption of a magnon, which provides both the required spin angular momentum and finite wavevector, activating otherwise forbidden optical features [Fig. 1(a)]. These transitions are commonly classified into cold and hot magnon sidebands, depending on whether magnons are created or annihilated \cite{MeltzerMnF2, TsuboiAhmet1992,Shinagawa1971JPSJ,Tanaka1971JPSJ,Fujiwara1972JPSJ}. 
The hot magnon sideband intensity is suppressed at low temperatures due to the reduced thermal magnon population  \cite{Shinagawa1971JPSJ,Tanaka1971JPSJ,Fujiwara1972JPSJ,Tsuboi1983PRB,Tsuboi1984PLA,TsuboiAhmet1992}. 
This naturally leads to the question of how a hot magnon sideband responds when a magnetic field closes the magnon gap or reconstructs the magnon spectrum across field-induced phases.
Although exciton-magnon sidebands have long been recognized  \cite{TanabeMoriyaSugano1965,Allen1966PRL,Sell1967,MeltzerMnF2,TsuboiAhmet1992,Shinagawa1971JPSJ,Tanaka1971JPSJ,Fujiwara1972JPSJ,Tsuboi1983PRB,Tsuboi1984PLA,TsuboiAhmet1992,Moore1981}, their response to external magnetic fields has remained largely unexplored. Understanding such field-induced evolution is crucial for controlling light–matter interactions via spin degrees of freedom.

\begin{figure}
    \centering
    \includegraphics[width=1\linewidth]{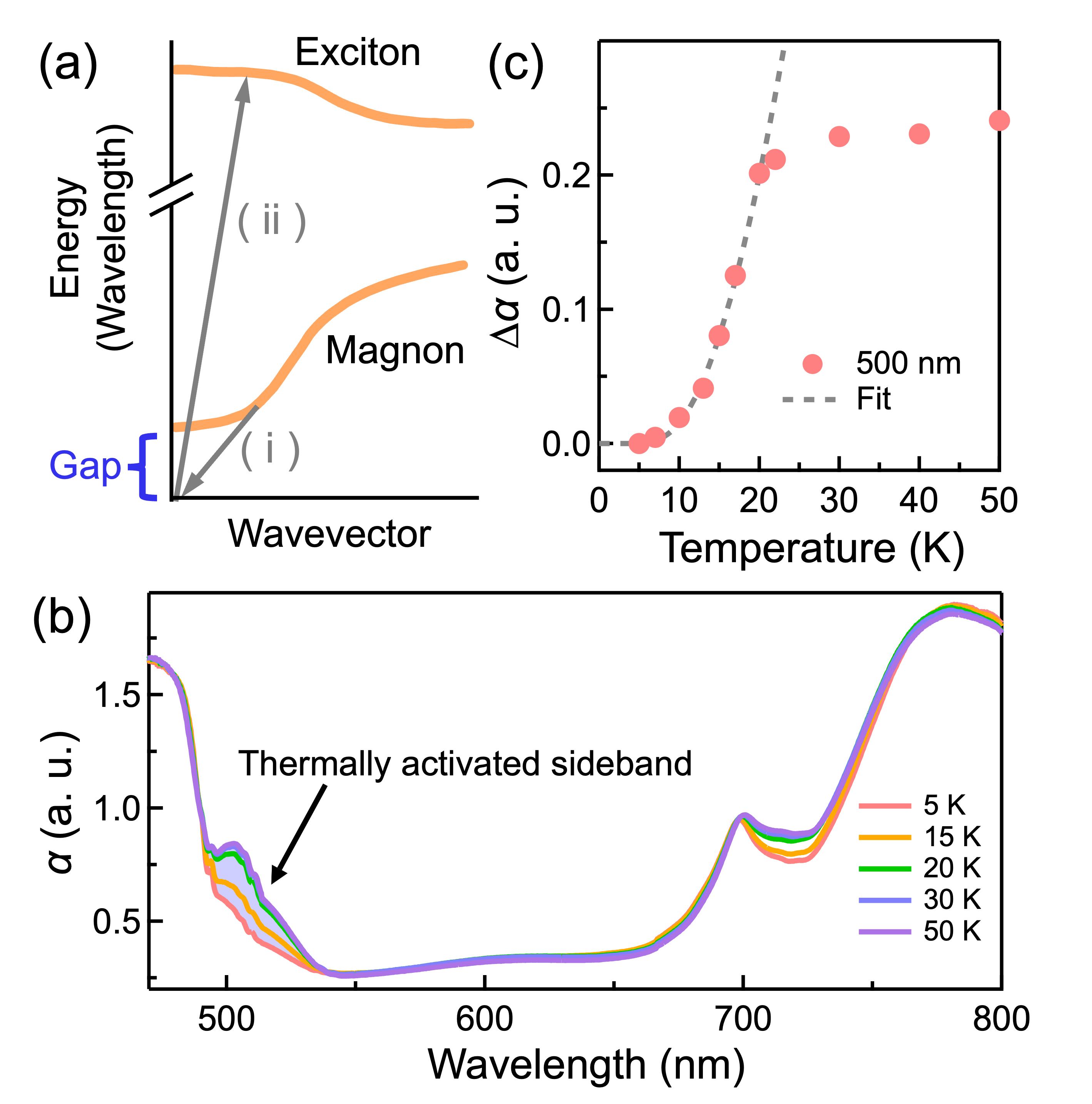}
    \caption{(a) Schematic illustration of the hot magnon sideband process.  The magnon dispersion has a finite energy gap. Annihilation of thermally populated magnons (i) is followed by exciton creation (ii). (b) Optical absorption spectra of LiNiPO$_4$ at selected temperatures in zero magnetic field using linearly polarized light ($\bm{E} \parallel a$). The arrow indicates the thermally activated sideband. (c) Temperature dependence of the hot-magnon sideband intensity of LiNiPO$_4$ at 500 nm. The dashed line shows a fit to a thermally activated model.}
    \label{fig:placeholder}
\end{figure}

LiNiPO$_4$, a representative magnetoelectric antiferromagnet, crystallizes in the orthorhombic space group \textit{Pnma} with strongly anisotropic lattice parameters, $a = 10.02\,\mathrm{\AA}$, $b = 5.83\,\mathrm{\AA}$, $c = 4.66\,\mathrm{\AA}$ \cite{Santoro1966_JPSCS,Mercier1967SSC,Abrahams1993,Vaknin2004}.
 Upon cooling, LiNiPO$_4$ undergoes a second-order transition into an intermediate incommensurate antiferromagnetic phase at $T_\text{N2}= 21.7$~K, followed by a sharp first-order transition into a commensurate antiferromagnetic phase at $T_\text{N1}= 20.8$~K \cite{Vaknin2004,Li2009PRB,ToftPetersen2011}. In the commensurate phase, the Ni$^{2+}$ moments  are predominantly aligned along the crystallographic $c$ axis with a small canting toward the $a$ axis \cite{Vaknin2004,Jensen2009PRB,ToftPetersen2011}. 
Such a complex spin configuration can give rise to unconventional magneto-optical phenomena \cite{Kimura2024PRL}. 
Importantly, the magnetic properties of LiNiPO$_4$ are governed by its strong magnetic anisotropy, Dzyaloshinskii–Moriya interactions, and competing exchange interactions beyond nearest neighbors, leading to a complex sequence of field-induced phases. Recently, high-field studies have revealed a rich magnetic phase diagram up to 56 T \cite{Fogh2020PRB}, accompanied by substantial modifications of the magnon excitation spectrum \cite{ToftPetersen2011}.
These properties make LiNiPO$_4$ an ideal platform for investigating magnetic-field control of exciton–magnon sideband processes.

In this study, we investigated both the temperature and magnetic-field dependence of magnon sideband intensities in the magnetoelectric antiferromagnet LiNiPO$_4$ using pulsed magnetic fields up to 50 T. The results reveal that the sideband intensity is sharply switched across magnetic phase transitions. Interestingly, plateau phases I and V exhibit strong suppression of the sideband intensity, accompanied by thermally activated behavior indicative of the presence of a finite magnon gap. In contrast, phases IV and VII show enhanced sideband intensity, reflecting their strongly canted spin structures. These results indicate that the magnetic-field control of spin degrees of freedom plays a key role in the selective switching of exciton-magnon optical transitions in antiferromagnets.

Single crystals of LiNiPO$_4$ were grown by the flux method \cite{Fomin2002}. A crystal was prepared as a thin plate ($\sim$50 {\textmu}m thick) with the surface parallel to the $c$ plane and mounted on a quartz substrate (2 mm in diameter and 0.3 mm thick).  Magneto-absorption spectra were measured in pulsed magnetic fields up to 50 T (pulse duration $\sim$36 ms) \cite{Miyata2026} in the Faraday configuration, with the magnetic field $\bm{B}$ and the light propagation direction $\bm{k}$ along the $c$ axis ($\bm{B} \parallel \bm{k} \parallel c$). The electric field of the incident light $\bm{E}$ was linearly polarized along the $a$ axis. Temperature-dependent measurements were performed between 4.2 K and 50 K in a liquid-helium cryostat. A broadband halogen lamp and a CCD-based spectrometer were used to record transmission spectra with an exposure time of 0.5 ms. Single-wavelength transmission measurements were also performed to monitor the magnetic-field dependence of the absorption intensity with higher temporal resolution as described in Ref. \cite{Bergsma2025RSI113903}.


\begin{figure}
    \centering
    \includegraphics[width=1\linewidth]{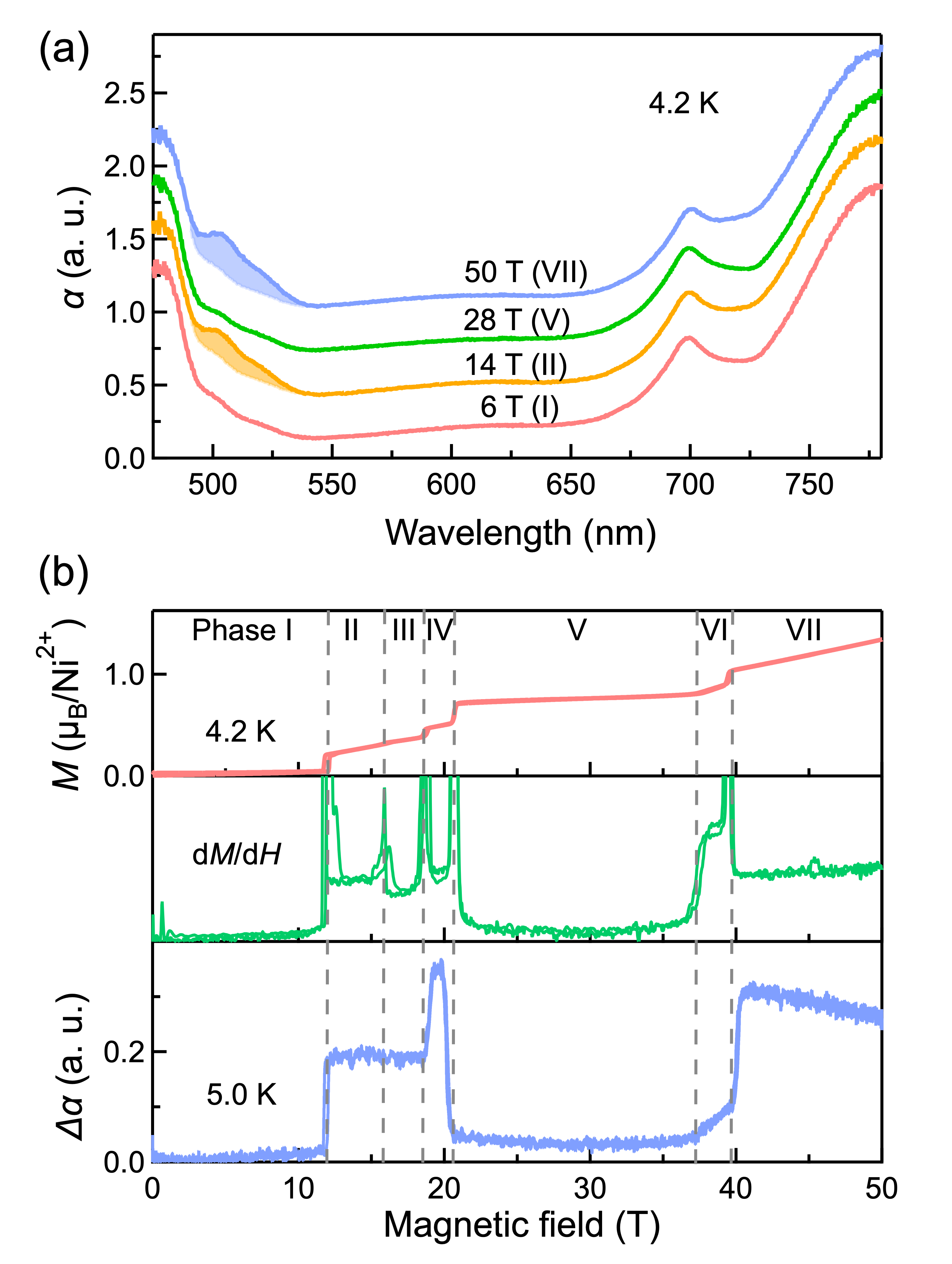}
    \caption{(a) Optical absorption spectra of LiNiPO$_4$ measured with linearly polarized light ($\bm{E} \parallel a$) for representative magnetic phases I (6 T), II (14 T), V (28 T), and VII (50 T) at 4.2 K. The shaded region highlights the magnon sideband, which appears and disappears by applying a magnetic field.
    (b) The magnetization $M$ of LiNiPO$_4$ from Ref. \cite{Fogh2020PRB} (upper panel), its differential susceptibility d$M$/d$H$ (middle panel), and the change in the optical absorption $\Delta\alpha$ at a wavelength of 500 nm and at 5 K (lower panel). 
    Vertical dashed lines indicate magnetic phase boundaries. The phase labels are taken from Ref. \cite{Fogh2020PRB}.}
    \label{fig:placeholder}
\end{figure}

Figure 1(b) shows the zero-field optical absorption spectra of LiNiPO$_4$ measured between 5 and 50 K for light polarized along the $a$ axis ($\bm{E} \parallel a$). A pronounced temperature dependence is observed around 500 nm, where the absorption is strongly suppressed at low temperatures and increases upon warming. 
To quantify this behavior, we plot the change in absorption, $\Delta \alpha$, relative to its value at 5 K, as shown in Fig. 1(c). This behavior is reminiscent of the typical hot magnon sideband, as observed in MnF$_2$ \cite{Shinagawa1971JPSJ, TsuboiAhmet1992}.
The temperature dependence follows an activated behavior and can be described by $\Delta \alpha(T) = A\exp\!\left(-\frac{E_{\mathrm g}}{k_{\mathrm B}T}\right)$, where $A$ is a prefactor, $T$ is the temperature, $k_{\mathrm B}$ is the Boltzmann constant, and $E_{\mathrm g}$ represents the activation energy. The fitting yields $E_{\mathrm g} = $4.7~meV, which is much larger than the reported magnon gap in LiNiPO$_4$ ($\sim$2.0~meV)  \cite{Vaknin2004,Jensen2009PRB}, suggesting that a simple Arrhenius form is insufficient. In practice, the prefactor $A$ acquires additional temperature dependence $\propto T^n$ reflecting the dimensionality $n$ of the magnon density of states \cite{Shinagawa1971JPSJ,Tanaka1971JPSJ,Fujiwara1972JPSJ, Tsuboi1984PLA,TsuboiAhmet1992}. 
A value of $n = 2$ gives the closest agreement with the reported magnon gap ($E_{\mathrm g} = $1.9 meV), consistent with quasi-two-dimensional magnetic behavior within the $bc$ planes observed in LiNiPO$_4$ \cite{Vaknin1999,Vaknin2004,Jensen2009PRB}. Note that the absorption around 720 nm also exhibits similar thermally activated behavior \footnote{The single-wavelength measurements at 730 nm show a magnetic-field dependence of the absorption changes similar to that observed in Fig. 2(b).}.

\begin{figure}
    \centering
    \includegraphics[width=1\linewidth]{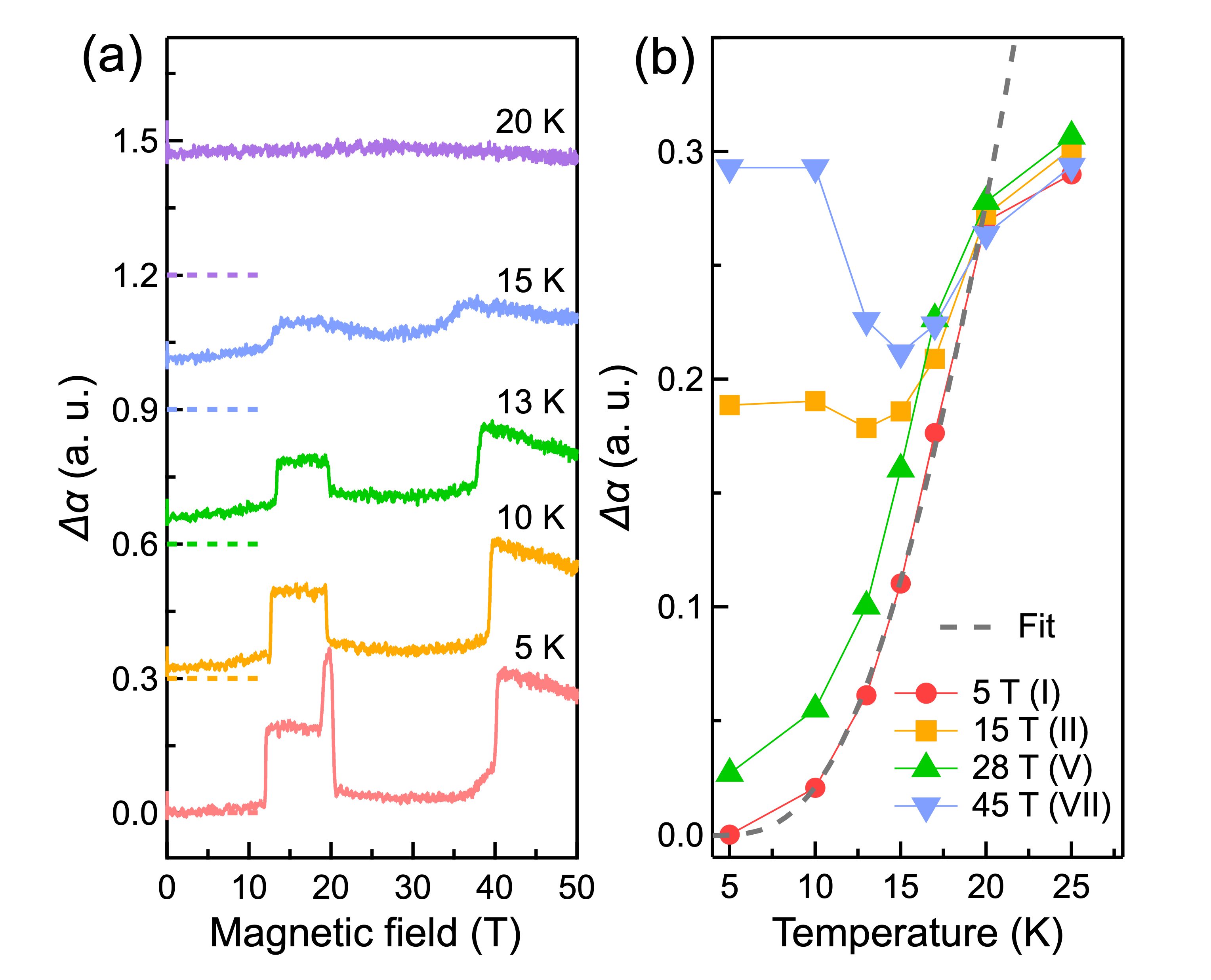}
    \caption{
    (a) Magnetic-field dependence of the change in optical absorption $\Delta\alpha$ at a wavelength of 500 nm measured at several temperatures with linearly polarized light ($\bm{E} \parallel a$). 
    The dashed lines indicate the zero level for each trace. 
    (b) Temperature dependence of the magnon sideband intensity extracted from $\Delta\alpha$ in Fig. 3(a) in representative magnetic phases I (5 T), II (15 T), V (28 T), and VII (45 T).  The dashed line shows a fit to a thermally activated model.
}
    \label{fig:placeholder}
\end{figure}

Previous optical studies on LiNiPO$_4$ \cite{Belletti1991PSSB,Kimura2024PRL} have identified several Ni$^\text{2+}$ $d–d$ transitions in the $C_\text{2v}$ environment.
However, spin-forbidden transitions from the $^{3}B_{2}$ ground state to low-lying singlet states ($^{1}A_{2}$, $^{1}B_{1}$, and $^{1}A_{1}$) have not been clearly resolved at their lowest temperatures. The observed sideband is attributed to these transitions activated through coupling to magnons \cite{Kimura2024PRL,Belletti1991PSSB}. 
The energy width of the observed sideband is on the order of 200 meV, which is much larger than the typical magnon bandwidth in antiferromagnets. This suggests that the width is not determined by the magnon dispersion of LiNiPO$_4$ but rather reflects the intrinsic linewidth of the electronic excitation.
Because these transitions involve different orbital characters, the electronic excitation can exhibit a typical intrinsic linewidth due to Franck–Condon effects \cite{Sugano1970}, contributing to the observed broad spectral feature. 
These orbital characters also lead to strongly anisotropic optical selection rules, with absorption predominantly observed for polarization along the $a$ axis (The data for polarization along the $b$ axis is shown in the Supplemental Material).

Magnetic fields strongly modify the magnon excitation spectrum across different magnetic phases, providing a tuning parameter for exciton–magnon coupling. To explore this effect, we investigate how the hot magnon sideband evolves under applied magnetic fields.
Figure 2(a) shows the magneto-absorption spectra measured at 4.2 K in representative magnetic phases I (6 T), II (14 T), V (28 T), and VII (50 T), using the phase labels described in Ref. \cite{Fogh2020PRB}. The magnon sideband exhibits a pronounced phase-dependent switching under applied magnetic fields. The hot magnon sideband is strongly suppressed in phases I and V, while it reappears in phases II and VII, as indicated by the shaded area in Fig. 2(a). 
The field-induced changes are most pronounced in the absorption feature around the wavelength of 500 nm, while the other 
$d–d$ excitations remain largely unaffected, indicating that the observed feature has a magnetic origin.

To track the field dependence more precisely, we perform single-wavelength transmission measurements at the sideband energy (wavelength of 500 nm). Figure 2(b) shows the field dependence of the sideband intensity, which exhibits sharp switching at the phase boundaries identified in the magnetization curve in Ref. \cite{Fogh2020PRB}. 
Interestingly, the intensity is strongly suppressed in phases I and V, while it is finite in the other phases. This behavior closely follows the differential magnetic susceptibility d$M$/d$H$. 
The strong suppression of the sideband intensity, together with the plateau-like behavior of the magnetization, suggests the presence of a finite magnon gap. Inelastic neutron scattering measurements at low magnetic fields have shown that the magnon gap in phase I softens at the transition to phase II \cite{ToftPetersen2011}, supporting that the observed switching of the sideband intensity around 12 T is consistent with the presence of a finite magnon gap.

The observed switching behavior can be understood in terms of two key factors: the thermal magnon population and the spin-dependent optical transition matrix element.
In general, the intensity of an optical transition can be expressed as $I(\omega) \propto |\mathcal{M}_\text{opt}|^2 D(\omega)$, where $\mathcal{M}_\text{opt}$ is the transition matrix element and $D(\omega)$ is the joint density of states (JDOS) \cite{HaugKoch1993,YuCardona2010}.
In the case of a magnon sideband, $\mathcal{M}_\text{opt}$ is strongly influenced by the spin configuration, while $D(\omega)$ is determined by the magnon spectrum and their thermal population \cite{Shinagawa1971JPSJ}.
The nearly vanishing intensity in phases I and V suggests that the thermal magnon population is effectively depleted,  
consistent with the presence of a finite magnon gap in these phases. 

Regarding the magnetic structures in LiNiPO$_4$, recent neutron studies \cite{Fogh2020PRB} have revealed that phase I is nearly collinear, phases II–III are elliptical spirals, phase V is a circular spiral (a collinear structure was also suggested \cite{ToftPetersen2017}), and phases IV and VII consist of two spins aligned along the $c$ axis and two strongly canted toward the $a$ axis. In phase VII, the spins gradually cant from the $a$ axis toward the $c$ axis with increasing magnetic field, accompanied by a reduction of the sideband intensity.
This observation suggests that the strongly canted spin structures in phases IV and VII, with large a-axis components, could enhance the transition matrix element $\mathcal{M}_\text{opt}$, thereby increasing the sideband intensity. 
This contrasts with the spiral phases (II, III), where the sideband intensity remains moderate despite their noncollinear spin structures.
A plausible explanation is that the spin flip associated with the local $d–d$ excitation (i.e., the exciton) couples to magnons through the spin Hamiltonian, particularly via antisymmetric exchange channels  \cite{Moriya1968}. In LiNiPO$_4$, $D_\text{DM}/J \sim 0.3$, where  $D_\text{DM}$ is the Dzyaloshinskii–Moriya interaction and  $J$ the main exchange interaction \cite{ToftPetersen2011,Peedu2019}, indicating that the antisymmetric interaction is sufficiently large to provide a key channel for exciton–magnon coupling.

To clarify whether the sideband intensity in each magnetic phase is mainly governed by $D(\omega)$, we investigate its thermally activated behavior in magnetic fields.
Since temperature-sweep measurements are not feasible within the millisecond duration of pulsed magnetic fields \cite{Miyata2026}, we perform single-wavelength transmission measurements at 500 nm while sweeping the magnetic field at several fixed temperatures, as shown in Fig. 3(a). The intensity is evaluated in terms of the change in absorption, $\Delta \alpha$, defined relative to its value at the lowest temperature and at zero magnetic field.
By comparing the sideband intensity at the same magnetic fields across these field sweeps, we reconstruct the temperature dependence at selected fields, as summarized in Fig. 3(b).

The sideband intensity in phases I and V exhibits an exponential-like increase with temperature, characteristic of a thermally activated process. This behavior closely resembles the typical zero-field hot magnon sideband \cite{Shinagawa1971JPSJ, TsuboiAhmet1992}. In contrast, the sideband intensity in phases II and VII shows a nonmonotonic temperature dependence, decreasing with increasing temperature and exhibiting a minimum around 15 K.
The nonmonotonic temperature dependence can be understood as a consequence of the distinct temperature evolution of the two contributing factors $\mathcal{M}_\text{opt}$ and $D(\omega)$. While $\mathcal{M}_\text{opt}$
is reduced by thermal fluctuations in the magnetic structure, $D(\omega)$
increases monotonically with temperature due to thermal excitations. The competition between these two effects gives rise to the observed temperature dependence.

\begin{figure}
    \centering
    \includegraphics[width=1\linewidth]{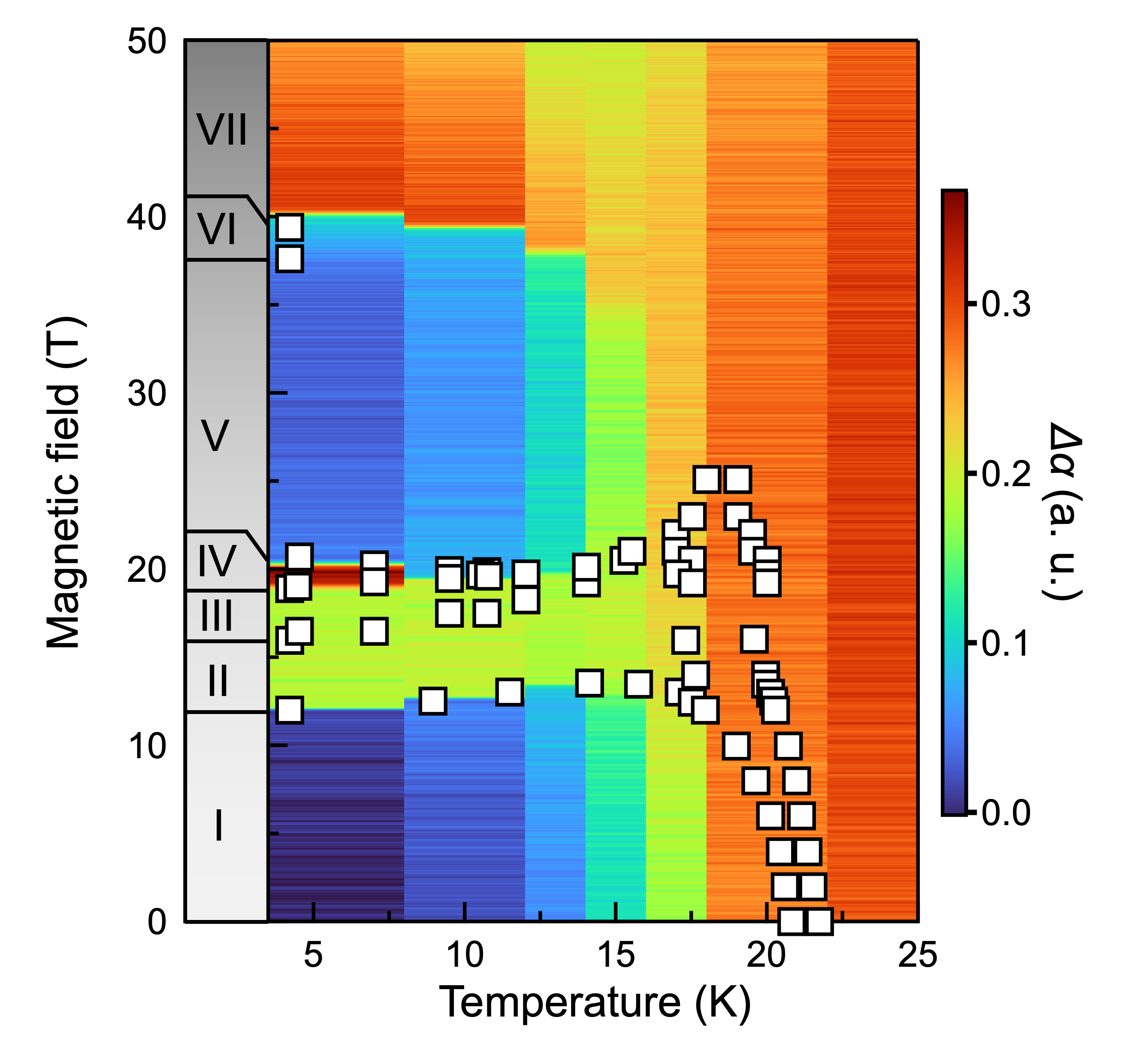}
    \caption{
    Field–temperature map of the optical absorption intensity at a wavelength of 500 nm, obtained from Fig. 3(a). The color scale represents $\Delta \alpha$. Open squares indicate magnetic phase boundaries shown in Ref. \cite{Fogh2020PRB}. The sideband intensity is strongly suppressed in phases I and V.
    }
    \label{fig:placeholder}
\end{figure}

Figure 4 shows the field–temperature map of the sideband intensity obtained from Fig. 3(a). 
The magnetic phase transition points, taken from Ref. \cite{Fogh2020PRB}, are indicated by open squares.
The optical intensity map shows good agreement with the magnetic phase diagram. The sideband intensity is strongly suppressed in phases I and V with thermal activation behavior, while it is enhanced in the other phases, exhibiting a clear on-off behavior across the field-induced magnetic phases. Notably, although phase IV is stable only in a narrow region of the phase diagram, it exhibits high intensity.
The optical intensity map directly captures the temperature-field evolution of the exciton-magnon coupling, establishing this optical approach as a sensitive and powerful probe of the magnetic phase diagram.


In summary, we investigated the magnon sideband intensity in LiNiPO$_4$ under high magnetic fields up to 50 T and found that exciton–magnon coupling is strongly controlled by both temperature and magnetic field. The magnon sideband intensity shows sharp switching, particularly with strong suppression in plateau phases I and V, indicating the presence of a finite magnon gap. In contrast, phases IV and VII show enhancement of the sideband intensity, reflecting their strongly canted spin structures. The switching is attributed to the interplay between the thermal magnon population
and the spin-dependent optical transition matrix element. Our results demonstrate that field-induced magnetic phases selectively enable or suppress exciton-magnon optical transitions, providing a new pathway toward controlling light–matter interactions in antiferromagnets.

%


\begin{acknowledgments}

 Z.Y. was supported by research grant from Research Foundation for Opto-Science and Technology and JSPS KAKENHI Grant Number 25K17324 (Grant-in-Aid for Early-Career Scientists). K. Kimura was partly supported by JSPS KAKENHI (Grant Nos. JP23K17663, JP24K00575, 26H00677, JP26H02194) and the Iketani Science and Technology Foundation. A.M. was supported by JSPS KAKENHI, Fund for the Promotion of Joint International Research (Home-Returning Researcher Development Research) No. 22K21359.

\end{acknowledgments}


\bibliography{Manuscript.bib}

\end{document}